\documentclass[
 reprint,
 amsmath,amssymb,
 aps,
 floatfix,
]{revtex4-2}
\usepackage{enumerate}
\usepackage{amsmath}
\usepackage{amsfonts}
\usepackage{amssymb}
\usepackage[utf8]{inputenc}
\usepackage[T1]{fontenc}
\usepackage{mathtools}
\usepackage{graphicx}
\usepackage[english]{babel}
\usepackage[svgnames]{xcolor}
\usepackage[colorlinks=true,citecolor=blue]{hyperref}
\usepackage{url}
\usepackage{float}
\usepackage{dcolumn}
\usepackage{bm}

\begin{document}

\preprint{SDANISO/2026}

\title{Localized magnetic pressure anisotropy in strange dwarfs with crystalline crusts}

\author{Mila Leite Garcia Reis}
\affiliation{Núcleo de Astrofísica e Cosmologia (Cosmo-Ufes) \& Departamento de Física, Universidade Federal do Espírito Santo, Espírito Santo, Brazil}

\author{Daniela Eller Uliana}
\affiliation{Núcleo de Astrofísica e Cosmologia (Cosmo-Ufes) \& Departamento de Física, Universidade Federal do Espírito Santo, Espírito Santo, Brazil}

\author{Jaziel G. Coelho}
\email{jaziel.coelho@ufes.br}
\affiliation{Núcleo de Astrofísica e Cosmologia (Cosmo-Ufes) \& Departamento de Física, Universidade Federal do Espírito Santo, Espírito Santo, Brazil}

\author{Edson Otoniel}
\email{edson.otoniel@ufca.edu.br}
\affiliation{Instítuto de Formação de Educadores , Universidade Federal do Carir\'i, R. Olegário Emidio de Araujo, s/n - Centro, Brejo Santo - CE, 63260-000, Brazil}

\date{\today}

\begin{abstract}
We study equilibrium sequences of magnetized strange dwarfs composed of a self bound MIT bag strange matter core, with \(B_{\rm bag}^{1/4}=145\,{\rm MeV}\), and an ordinary crystalline crust. The aim is to determine how nuclear composition changes and magnetic pressure anisotropy in the crust modify the mass radius relation. The strange core is kept isotropic, while the crust is described either by pure nuclei stopped at the first inverse beta threshold or by electron capture sequences extended to neutron drip for C, O, Ne, and Mg compositions; selected mixed crusts are also considered up to the first instability. The stellar structure is solved with the radial pressure as the integrated pressure and with the anisotropic contribution \(2(P_t-P_r)/r\) restricted to the crust. This treatment is compared with ordinary white dwarf sequences and with scalar pressure branch calculations. We find that extending the crust to neutron drip produces more compact strange dwarf branches than stopping at the inverse beta threshold, because the evolved crust is softer and the core crust transition occurs at higher pressure. Magnetic anisotropy further shifts the selected branches toward smaller radii, with the effect visible across the compositions studied and clearer at fixed stellar mass. The comparison with compact objects from the Montreal White Dwarf Database is used only as an observational reference plane, but it indicates that compact white dwarf candidates are a relevant region for testing small core strange dwarf scenarios.
\end{abstract}

\maketitle

\section{Introduction}
\label{sec:introduction}

The possible existence of compact objects containing strange quark matter is a direct consequence of the hypothesis that matter with approximately equal fractions of up, down, and strange quarks may be more stable than ordinary nuclear matter at sufficiently high density. Within this scenario, strange matter can form self bound stellar cores whose equilibrium is not determined only by gravity, but also by the microscopic stability of the quark phase. Strange dwarfs represent a particular realization of this idea. They are hybrid compact configurations composed of a self bound strange quark matter core surrounded by an extended crust of ordinary matter, with the two regions separated by an electrostatic barrier that supports the nuclear crust against direct conversion into strange matter. The original construction showed that equilibrium sequences with strange matter cores and nuclear crusts can extend continuously from compact strange stars to white dwarf like configurations, thereby defining a class of dense dwarfs with macroscopic properties different from ordinary white dwarfs \cite{Glendenning_1995,Glendenning_1995_ApJ}. Subsequent studies have returned to this problem in connection with candidate compact white dwarfs, strange matter cores, stability criteria, and the observational degeneracy between strange dwarfs and conventional white dwarfs \cite{Mathews_2006,Di_Clemente_2024,da_Silva_2026}.

The defining structural feature of a strange dwarf is the coexistence of two physically distinct domains. The inner core is made of dense self bound strange matter, whereas the external layers are formed by ordinary nuclei and electrons. Since the crust is extended over a much larger radial scale than the core, it can dominate the stellar radius even when the strange core contributes substantially to the gravitational mass. This property makes strange dwarfs qualitatively different from neutron stars, whose radii are closely connected to the equation of state at supranuclear density, and also from ordinary white dwarfs, whose entire support is provided by electron degeneracy pressure in ordinary matter. In strange dwarfs the dense core fixes the inner boundary condition and contributes to the compactness, while the crust determines where the surface is reached. The mass radius relation is therefore especially sensitive to the crust equation of state, to the density at the base of the crust, and to the matching prescription at the interface with strange matter \cite{Glendenning_1995,Di_Clemente_2024}. This sensitivity also explains why strange dwarfs may be difficult to distinguish from white dwarfs through mass and radius measurements alone, particularly when rotation or composition effects are included \cite{da_Silva_2026}.

The observational motivation follows from the same structural property. If a white dwarf contains a strange matter core, its radius at fixed mass can be smaller than that of a conventional electron degenerate configuration, producing a compact branch that may overlap with part of the observed white dwarf population \cite{Mathews_2006}. This possibility has been used to motivate searches for objects whose inferred radii are difficult to reconcile with standard white dwarf models, although such identifications remain dependent on uncertainties in mass, radius, atmosphere modeling, composition, and evolutionary history. Theoretical studies of strange dwarfs must therefore separate robust structural effects from model artifacts. Since the crust is the region most directly connected with the radius, any modification of the crustal stress tensor is amplified in the observable part of the mass radius diagram.

The core crust interface is not a secondary modeling detail. The stability of strange dwarfs depends on the boundary conditions imposed between the nuclear crust and the quark matter core, as emphasized by modern analyses of the apparent tension between different radial stability studies \cite{Di_Clemente_2024}. The existence of an electrostatically suspended crust allows ordinary matter to remain outside the quark phase, but the global stellar model still requires a consistent treatment of pressure, energy density, gravitational mass, and the matching radius. Since the crust controls a large part of the radius, any physical mechanism that changes the pressure support in this region can produce a measurable displacement in the mass radius curve. This observation is central for magnetized strange dwarfs, because the magnetic response of the ordinary crust need not be the same as that of the self bound strange core. A physically meaningful treatment must therefore identify where the magnetic stress is generated and how it is coupled to hydrostatic balance.

Strong magnetic fields are a natural ingredient of compact star physics. Neutron stars typically have surface fields of order \(10^{12}\) to \(10^{13}\,{\rm G}\), while magnetars reach values near \(10^{14}\) to \(10^{15}\,{\rm G}\), and their internal fields may be substantially larger \cite{Lander_2009,Ioka_2004}. White dwarfs also provide an important context for magnetic structure calculations. Magnetic white dwarfs span a broad observational range of field strengths and appear both as isolated objects and in binary systems \cite{Ferrario2015,Kepler2013,Landstreet2019,Bagnulo2022}. In interacting systems, pulsed radio, optical, X-ray, and polarimetric emission show that magnetic white dwarfs can also act as compact engines in binaries \cite{Marsh2016,Buckley2017,Pelisoli2023,otoniel_white_2017,malheiro_double_2026}. Cyclotron features and magnetically driven outflows further indicate that the field can affect radiation and angular momentum loss in white dwarf systems \cite{Schmidt2001,Kashiyama2019}. Magnetic white dwarfs can possess surface fields far below the extreme magnetar scale, but theoretical models allow much stronger internal fields, and self consistent calculations have shown that magnetic stresses, general relativity, rotation, field geometry, deformation, and corrections to the equation of state can alter the mass radius relation \cite{Jordan_2003,Franzon_2015,Bera_2016,otoniel_effects_2017,otoniel_strongly_2019,sousa_prospects_2024}. These results show that the magnetic field cannot always be regarded as a passive label attached to an otherwise isotropic equation of state. Its effect depends on the microscopic response of charged matter, on the geometry of the field, and on the way the electromagnetic stress enters the gravitational equilibrium problem.

A magnetic field introduces a preferred spatial direction and therefore changes the tensor structure of the stress energy tensor. In a magnetized Fermi system, the pressure transverse to the magnetic field is not generally equal to the pressure parallel to it. The relation between the parallel and transverse components involves the magnetization, and the degree of anisotropy increases with the magnetic field strength in the relevant regimes \cite{Canuto1968,Perez2008,Strickland_2012}. The same conceptual issue appears in magnetized quark matter, where the equation of state may provide different longitudinal and transverse pressures, and where the use of a single pressure in spherical stellar equations must be interpreted with care \cite{Menezes_2015}. Magnetic pressure anisotropy is therefore not a phenomenological scalar correction. It is a tensorial property associated with the directional character of the magnetic field and with the decomposition of the stress energy tensor into components parallel and perpendicular to that direction.

This distinction is important because the usual Tolman Oppenheimer Volkoff formalism assumes an isotropic fluid. If the equation of state supplies different radial and tangential stresses, replacing the pressure by one scalar branch can be useful as a diagnostic calculation, but it does not represent the complete anisotropic force balance. Studies of magnetized white dwarfs and anisotropic relativistic stars have shown that anisotropic stresses can modify equilibrium sequences, field structure, and the domain of regular solutions \cite{Alvear_Terrero_2015,Folomeev_2015,Horvat2011,Raposo2019}. In fully magnetized stellar models, the magnetic field also contributes to the spacetime through the Maxwell energy momentum tensor and to the matter dynamics through Lorentz forces and magnetization effects \cite{Franzon_2015,Bera_2016,Paret2015,Franzon2016}. A spherical anisotropic treatment is therefore an approximation to a more general axisymmetric problem, but it remains a controlled way to isolate the local consequence of unequal radial and tangential pressures in hydrostatic balance.

Many studies of anisotropic compact stars assume that the anisotropy acts throughout the whole stellar interior. That assumption is adequate for exploring generic effects of anisotropic matter or for modeling objects in which the microscopic origin of anisotropy is distributed over the dense phase. Strange dwarfs require a different separation of physical domains. Their strange matter core is self bound and, in the model considered here, is expected to remain approximately isotropic with respect to the crustal magnetic pressure splitting. The natural location of magnetic pressure anisotropy is the ordinary crust, where charged nuclei and electrons interact directly with the magnetic field. Extending the crustal anisotropic stress into the strange core would mix two different equations of state and would artificially modify the region that is not responsible for the crustal magnetic response. The relevant question is thus not only whether \(P_r\) and \(P_t\) differ in the magnetized crust, but whether the stellar structure changes when \(P_t-P_r\) is allowed to act only where the crustal equation of state is physically valid.

This localization is particularly relevant for radius predictions. Since the crust carries most of the radial extension of a strange dwarf, an anisotropic force confined to the crust can shift the surface radius even if the strange core is left unchanged. Conversely, applying the same anisotropic prescription over the entire star can produce changes in mass and compactness that do not originate from the magnetized ordinary matter layer. The resulting mass radius curve then reflects a mixture of crust physics and an imposed core modification. For this reason, magnetized strange dwarfs provide a useful theoretical setting in which to separate scalar pressure effects from the actual anisotropic force term. The comparison is also important observationally, because strange dwarfs may mimic white dwarfs in portions of the mass radius plane, and small changes in predicted radii can affect the interpretation of compact white dwarf candidates \cite{Mathews_2006,da_Silva_2026}.

The remainder of this article is organized as follows. Section~\ref{sec:eos} defines the strange matter core, the crust EOS, the lattice correction, and the anisotropic stress tensor used to construct the radial and tangential pressure branches. Section~\ref{sec:structure} presents the stellar structure equations and the coupling between an isotropic core and an anisotropic crust. Section~\ref{sec:results} discusses scalar pressure diagnostics and localized anisotropic sequences for different crust configurations. Section~\ref{sec:conclusions} summarizes the physical implications and limitations of the model.

\section{Equation of State and Anisotropic Stress Tensor}
\label{sec:eos}

The stellar configurations considered here are described by a self bound strange quark matter core surrounded by an ordinary matter crust. The local energy density entering the stellar structure equations is denoted by \(\epsilon\). The pressure integrated in the radial hydrostatic equation is denoted by \(P_r\), and the pressure acting in the angular directions is denoted by \(P_t\). In an isotropic fluid one has \(P_r=P_t=P\), where \(P\) is a scalar pressure. In a magnetized crust this equality is not generally preserved, because the magnetic field selects a preferred direction in the local stress tensor.

For the core we use a self bound MIT bag type equation of state, as commonly adopted in strange dwarf models with a quark core and a white dwarf like envelope \cite{Glendenning_1995,Glendenning_1995_ApJ,Di_Clemente_2024,da_Silva_2026}. The quark flavors are labeled by \(f=u,d,s\), corresponding to up, down, and strange quarks. At zero temperature, and in units \(\hbar=c=1\), the chemical potential of each flavor is
\begin{equation}
\mu_f=\left(k_{Ff}^2+m_f^2\right)^{1/2},
\label{eq:quark_mu}
\end{equation}
where \(k_{Ff}\) is the Fermi momentum and \(m_f\) is the rest mass of flavor \(f\). The number density is
\begin{equation}
n_f=\frac{k_{Ff}^3}{\pi^2},
\label{eq:quark_density}
\end{equation}
where spin and color degeneracies have been included. The pressure of the quark phase is
\begin{equation}
P_{\rm q}=-B_{\rm bag}+\sum_{f=u,d,s}P_f,
\label{eq:core_pressure}
\end{equation}
where \(B_{\rm bag}\) is the bag constant and \(P_f\) is the kinetic contribution of flavor \(f\), given by
\begin{equation}
P_f=\frac{1}{4\pi^2}\left[
\mu_f k_{Ff}\left(\mu_f^2-\frac{5}{2}m_f^2\right)
+\frac{3}{2}m_f^4
\ln\left(\frac{\mu_f+k_{Ff}}{m_f}\right)
\right].
\label{eq:quark_flavor_pressure}
\end{equation}
The corresponding energy density is
\begin{equation}
\epsilon_{\rm q}=B_{\rm bag}+\sum_{f=u,d,s}\epsilon_f,
\label{eq:core_energy}
\end{equation}
with
\begin{equation}
\epsilon_f=\frac{1}{4\pi^2}\left[
\mu_f k_{Ff}\left(\mu_f^2-\frac{1}{2}m_f^2\right)
-\frac{1}{2}m_f^4
\ln\left(\frac{\mu_f+k_{Ff}}{m_f}\right)
\right].
\label{eq:quark_flavor_energy}
\end{equation}
The core is treated as isotropic in this work. Thus,
\begin{equation}
P_r=P_t=P_{\rm q},
\qquad
\epsilon=\epsilon_{\rm q},
\label{eq:core_isotropic}
\end{equation}
inside the strange matter region. This assumption isolates the magnetic anisotropy of the ordinary crust and prevents the crustal magnetic stress from being extended into the self bound quark phase.

The crust is composed of fully ionized nuclei and degenerate electrons. To include both pure and mixed layers in the same notation, we label the nuclear species by \(j=1,2\). Each species is specified by its charge number \(Z_j\), baryon number \(A_j\), ion number density \(n_j\), and nuclear mass \(M_j=M(A_j,Z_j)\). Charge neutrality gives
\begin{equation}
n_e=\sum_{j=1}^{2}Z_j n_j,
\label{eq:charge_neutrality}
\end{equation}
where \(n_e\) is the electron number density. The total ion density, ionic fractions, and average charge and baryon numbers are
\begin{equation}
\begin{aligned}
n_I=\sum_{j=1}^{2}n_j,
\qquad
x_j=\frac{n_j}{n_I},\\
\bar Z=\sum_{j=1}^{2}x_jZ_j,
\qquad
\bar A=\sum_{j=1}^{2}x_jA_j .
\end{aligned}
\label{eq:ion_fractions}
\end{equation}
The baryon number density and electron fraction are then
\begin{equation}
n_B=\sum_{j=1}^{2}A_j n_j=n_I\bar A,
\qquad
Y_e=\frac{n_e}{n_B}=\frac{\bar Z}{\bar A}.
\label{eq:baryon_density_electron_fraction}
\end{equation}
The ionic rest energy density is
\begin{equation}
\epsilon_N=c^2\sum_{j=1}^{2}n_j M_j,
\label{eq:nuclear_energy_density}
\end{equation}
where \(c\) is the speed of light. Equivalently, the ionic mass density is
\begin{equation}
\rho_{\rm ion}=\sum_{j=1}^{2}n_j M_j ,
\label{eq:ion_mass_density}
\end{equation}
where the experimental nuclear masses \(M_j\) are retained. If they are approximated by \(A_jm_u\), with \(m_u\) the atomic mass unit, then \(\rho_{\rm ion}\simeq m_u n_B\). The one component limit is recovered by setting one ionic fraction to zero. For example, if \(x_1=1\), then \(n_e=Zn_i\), \(n_B=An_i\), \(Y_e=Z/A\), \(\epsilon_N=n_iM(A,Z)c^2\), and \(\rho_{\rm ion}=n_iM(A,Z)\).

Before introducing the magnetic field, the zero temperature crust equation of state can be written as
\begin{equation}
\epsilon_{\rm crust}^{(0)}
=\epsilon_N+\epsilon_e^{(0)}+\epsilon_{\rm L},
\label{eq:crust_energy_zero_field}
\end{equation}
and
\begin{equation}
P_{\rm crust}^{(0)}
=P_e^{(0)}+P_{\rm L}.
\label{eq:crust_pressure_zero_field}
\end{equation}
Here \(\epsilon_{\rm crust}^{(0)}\) and \(P_{\rm crust}^{(0)}\) are the nonmagnetized crust energy density and pressure, \(\epsilon_e^{(0)}\) and \(P_e^{(0)}\) are the usual zero field degenerate electron energy density and pressure, and \(\epsilon_{\rm L}\) and \(P_{\rm L}\) are the Coulomb lattice contributions. The cold ion pressure is neglected in Eq.~(\ref{eq:crust_pressure_zero_field}), while the ionic rest energy is kept in Eq.~(\ref{eq:crust_energy_zero_field}).

The lattice is taken to be body centered cubic. For the binary mixtures used in the EOS tables, the Coulomb contribution is written in a linear mixing form,
\begin{equation}
\epsilon_{\rm L}
=C_{\rm bcc} e^2 n_e^{4/3}
\frac{x_1Z_1^{5/3}+x_2Z_2^{5/3}}{\bar Z},
\label{eq:lattice_energy_binary}
\end{equation}
where \(e\) is the elementary charge and \(C_{\rm bcc}=-1.444231\) is the body centered cubic lattice constant in the convention used for Coulomb crystals \cite{Chamel_2014}. If one component is absent, Eq.~(\ref{eq:lattice_energy_binary}) reduces to the usual one component expression
\begin{equation}
\epsilon_{\rm L}=C_{\rm bcc} e^2 n_e^{4/3} Z^{2/3}
=C_{\rm bcc} e^2 Z^2 n_i^{4/3}.
\label{eq:lattice_energy_one_component}
\end{equation}

The lattice pressure is
\begin{equation}
P_{\rm L}=\frac{1}{3}\epsilon_{\rm L}.
\label{eq:lattice_pressure}
\end{equation}
Since \(C_{\rm bcc}<0\), the BCC lattice lowers the energy density and contributes a negative pressure correction.

The magnetic field first enters the crust through the electron sector. We define the electron critical field
\begin{equation}
B_c=\frac{m_e^2c^3}{e\hbar}
\simeq 4.414\times10^{13}\,{\rm G},
\label{eq:critical_field}
\end{equation}
where \(m_e\) is the electron mass and \(\hbar\) is the reduced Planck constant. The dimensionless magnetic field is
\begin{equation}
B_\star=\frac{B}{B_c},
\label{eq:b_star}
\end{equation}
where \(B\) is the magnetic field strength. The electron Compton wavelength is
\begin{equation}
\lambda_e=\frac{\hbar}{m_e c}.
\label{eq:electron_compton}
\end{equation}
For a uniform magnetic field, the electron Landau levels are
\begin{equation}
E_{\nu}(p_z)=
\left[
c^2p_z^2+m_e^2c^4\left(1+2\nu B_\star\right)
\right]^{1/2},
\label{eq:landau_energy}
\end{equation}
where \(E_{\nu}\) is the electron energy, \(p_z\) is the momentum parallel to the field, and \(\nu=0,1,2,\ldots\) labels the Landau level. The spin degeneracy is
\begin{equation}
g_\nu=
\begin{cases}
1, & \nu=0,\\
2, & \nu\ge 1.
\end{cases}
\label{eq:landau_degeneracy}
\end{equation}
If \(\mu_e\) is the electron chemical potential and \(\gamma_e=\mu_e/(m_ec^2)\), the dimensionless Fermi momentum in the level \(\nu\) is
\begin{equation}
x_e(\nu)=\left(\gamma_e^2-1-2\nu B_\star\right)^{1/2}.
\label{eq:xe_nu}
\end{equation}
Only levels for which \(x_e(\nu)\) is real are occupied, so that
\begin{equation}
\nu_{\rm max}\le
\frac{\gamma_e^2-1}{2B_\star}
<\nu_{\rm max}+1,
\label{eq:nu_max}
\end{equation}
where \(\nu_{\rm max}\) is the largest nonnegative integer satisfying the occupation condition. The electron number density is then
\begin{equation}
n_e=
\frac{2B_\star}{(2\pi)^2\lambda_e^3}
\sum_{\nu=0}^{\nu_{\rm max}}g_\nu x_e(\nu).
\label{eq:electron_density_landau}
\end{equation}

The electron energy density is
\begin{equation}
\epsilon_e(B)=
\frac{B_\star m_e c^2}{(2\pi)^2\lambda_e^3}
\sum_{\nu=0}^{\nu_{\rm max}}
g_\nu\left(1+2\nu B_\star\right)
\psi_+
\left[
\frac{x_e(\nu)}
{\left(1+2\nu B_\star\right)^{1/2}}
\right],
\label{eq:electron_energy_landau}
\end{equation}
and the radial electron pressure branch used in the effective spherical model is
\begin{equation}
P_{e,r}(B)=
\frac{B_\star m_e c^2}{(2\pi)^2\lambda_e^3}
\sum_{\nu=0}^{\nu_{\rm max}}
g_\nu\left(1+2\nu B_\star\right)
\psi_-
\left[
\frac{x_e(\nu)}
{\left(1+2\nu B_\star\right)^{1/2}}
\right].
\label{eq:electron_radial_pressure}
\end{equation}
The auxiliary functions appearing in Eqs.~(\ref{eq:electron_energy_landau}) and (\ref{eq:electron_radial_pressure}) are
\begin{equation}
\psi_{\pm}(x)=
x\left(1+x^2\right)^{1/2}
\pm
\ln\left[x+\left(1+x^2\right)^{1/2}\right],
\label{eq:psi_pm}
\end{equation}
where \(x\) is a dimensionless momentum argument. Equations~(\ref{eq:electron_density_landau}), (\ref{eq:electron_energy_landau}), and (\ref{eq:electron_radial_pressure}) are the standard zero temperature Landau level expressions used in magnetized white dwarf matter \cite{Chamel_2014,otoniel_fermionic_2015}. The magnetic field selects a preferred microscopic direction, which is represented in the spherical stellar model by the radial pressure branch \(P_{e,r}\).

The magnetization correction determines the tangential electron branch,
\begin{equation}
{\cal M}_e=
\left(\frac{\partial P_{e,r}}{\partial B}\right)_{\mu_e},
\label{eq:electron_magnetization}
\end{equation}
where \({\cal M}_e\) is the electron magnetization at fixed chemical potential. The tangential electron pressure is
\begin{equation}
P_{e,t}(B)=P_{e,r}(B)-{\cal M}_e B.
\label{eq:electron_tangential_pressure}
\end{equation}
This relation is obtained from the matter part of the energy momentum tensor for magnetized fermions \cite{Strickland_2012}. It is the microscopic origin of the pressure splitting in the crust: the magnetic field fixes a preferred direction and the tangential compression contains the magnetization work term.

The crust energy density and pressure branches used in the present EOS are then
\begin{equation}
\epsilon_{\rm crust}(B)
=\epsilon_N+\epsilon_e(B)+\epsilon_{\rm L},
\label{eq:energy_corrected}
\end{equation}
\begin{equation}
P_r(B)
=P_{e,r}(B)+P_{\rm L},
\label{eq:pr_total}
\end{equation}
and
\begin{equation}
P_t(B)
=P_{e,t}(B)+P_{\rm L}
=P_{e,r}(B)-{\cal M}_eB+P_{\rm L}.
\label{eq:pt_total}
\end{equation}
The cold ion pressure has again been neglected. We solve the stellar equilibrium numerically using \(\epsilon_{\rm crust}\), \(P_r\), and \(P_t\) as the matter variables. %No separate pure Maxwell contribution is added to the energy density or to the pressure branches in this implementation. 
f In this implementation, the pure Maxwell contribution ($B^2/8\pi$) is not added as a global scalar source. This allows the model to focus exclusively on the microscopic matter response (specifically the magnetization) and how these internal stresses alter hydrostatic balance without the model-dependent assumptions of a specific global magnetic field geometry \cite{Franzon_2015}.

The anisotropy parameter is defined by
\begin{equation}
\Delta(\epsilon,B)=P_t(\epsilon,B)-P_r(\epsilon,B).
\label{eq:delta}
\end{equation}
Using Eqs.~(\ref{eq:pr_total}) and (\ref{eq:pt_total}), one obtains
\begin{equation}
\Delta(\epsilon,B)
=P_t-P_r
=-{\cal M}_eB.
\label{eq:delta_magnetic}
\end{equation}
For \(B=0\), the magnetization contribution vanishes and the isotropic limit \(P_r=P_t\) is recovered.

The core crust matching is imposed at a pressure \(P_{\rm join}\). The matching radius \(R_{\rm core}\) is defined by
\begin{equation}
P_r(R_{\rm core})=P_{\rm join},
\label{eq:matching_pressure}
\end{equation}
where \(R_{\rm core}\) is the radius of the strange matter core. The pressure is continuous at the interface,
\begin{equation}
P_{\rm q}(R_{\rm core}^-)=P_r(R_{\rm core}^+)=P_{\rm join},
\label{eq:pressure_continuity}
\end{equation}
where \(R_{\rm core}^-\) and \(R_{\rm core}^+\) denote the inner and outer sides of the interface. The energy density may be discontinuous because self bound strange matter is matched to ordinary crust matter. In the present calculation the anisotropy satisfies \(\Delta=0\) in the core and \(\Delta\ne0\) only in the magnetized crust, where the ordinary charged matter shares the same microphysical ingredients used in magnetized white dwarf models \cite{Chamel_2014,Alvear_Terrero_2015,Franzon_2015,Bera_2016}. This construction extends magnetic anisotropic formalisms used in neutron star and white dwarf studies to strange dwarfs while preserving the isotropic character of the strange matter core \cite{Strickland_2012,Lander_2009,Ioka_2004,Menezes_2015,Folomeev_2015}. The quantities \(\epsilon_{\rm crust}\), \(P_r\), \(P_t\), and \(\Delta\) defined above are the inputs for the anisotropic stellar structure equations introduced next.

\subsection{Electron capture and neutron drip thresholds}
\label{subsec:capture_drip}

The crust composition is limited by weak interaction thresholds at high density. We treat inverse beta decay as a local thermodynamic instability, not as a reaction rate. For a parent nucleus \((A,Z)\), the next allowed capture in the prescribed sequence is tested by comparing the Gibbs free energy per baryon of the parent and daughter states at the same pressure, following the standard dense crust construction \cite{Baym1971BPS,Chamel_2014}. The nuclear masses entering this comparison are experimental masses when available, consistently with the mass convention used in the EOS construction \cite{Wang2021AME}. In the notation of this work, the relevant quantity is \(g=(\epsilon+P_r)/n_B\). The daughter density is chosen so that its radial pressure equals the parent pressure, and the capture threshold is the point where \(g_{\rm daughter}-g_{\rm parent}=0\). Below this point the parent layer is kept unchanged; above it the next nucleus in the sequence defines the deeper layer.

For the pure composition sequences, the chains keep the mass number fixed and lower the charge according to the selected capture channel. Thus carbon, oxygen, neon, and magnesium layers are followed through a finite set of nuclei until either the next capture occurs or neutron drip is reached. The calculation is therefore a restricted crust sequence, not a minimization over all possible nuclei. For binary mixtures, the same criterion is applied to each component, and the mixed branch is stopped at the first inverse beta instability reached by either species. The mixed configurations used later should therefore be read as fixed fraction models before capture, not as complete weak equilibrium mixtures.

The crust description without free neutrons ends at neutron drip. We locate this limit by testing the onset of a channel in which an electron capture is accompanied by the appearance of a free neutron, \((A,Z)+e^-\rightarrow(A-1,Z-1)+n+\nu_e\). At the threshold, the state after drip is pressure matched to the parent state and includes the neutron rest mass contribution, while the pressure of the free neutrons is neglected at onset. When the corresponding Gibbs difference vanishes, the EOS is terminated because matter beyond that point requires free neutron degrees of freedom that are not included in the present crust model.

\section{Structure of anisotropic magnetic strange dwarfs}
\label{sec:structure}

The stellar configurations considered here are static, spherically symmetric effective models in which the strange quark matter core and the ordinary matter crust are treated differently. The core is self bound and is not assumed to carry a magnetic pressure anisotropy in the present calculation. The crust, by contrast, contains charged ordinary matter whose equation of state provides two pressure branches, \(P_r\) and \(P_t\), at the same local energy density. The structure problem must therefore preserve the usual relativistic gravitational balance while allowing the anisotropic force to act only where the magnetized crust is present.

The spacetime is written in Schwarzschild like coordinates as
\begin{equation}
ds^2
=e^{2\Phi(r)}c^2dt^2
-e^{2\Lambda(r)}dr^2
-r^2d\Omega^2,
\label{eq:static_metric}
\end{equation}
where \(\Phi(r)\) and \(\Lambda(r)\) are metric functions and \(d\Omega^2=d\theta^2+\sin^2\theta\,d\phi^2\). The radial metric component is expressed in terms of the enclosed gravitational mass \(m(r)\) as
\begin{equation}
e^{-2\Lambda(r)}
=1-\frac{2Gm(r)}{rc^2},
\label{eq:lambda_mass}
\end{equation}
where \(G\) is Newton's constant and \(c\) is the speed of light.

For an anisotropic fluid, the stress energy tensor can be written in covariant form as
\begin{equation}
T_{\mu\nu}
=
\left(\epsilon+P_t\right)u_\mu u_\nu
-P_t g_{\mu\nu}
+\left(P_r-P_t\right)s_\mu s_\nu,
\label{eq:covariant_anisotropic_tensor}
\end{equation}
where \(u^\mu\) is the fluid four velocity, \(s^\mu\) is the unit radial spacelike vector, \(\epsilon\) is the local energy density, \(P_r\) is the radial pressure, and \(P_t\) is the tangential pressure. These vectors satisfy
\begin{equation}
u^\mu u_\mu=1,
\qquad
s^\mu s_\mu=-1,
\qquad
u^\mu s_\mu=0.
\label{eq:us_normalization}
\end{equation}
In the comoving frame one has \(u^\mu=(e^{-\Phi},0,0,0)\) and \(s^\mu=(0,e^{-\Lambda},0,0)\), and Eq.~(\ref{eq:covariant_anisotropic_tensor}) becomes
\begin{equation}
T^\mu{}_\nu
=
{\rm diag}
\left(
\epsilon,
-P_r,
-P_t,
-P_t
\right),
\label{eq:structure_tensor}
\end{equation}
which is the local effective tensor structure used for the anisotropic crust. In the isotropic limit \(P_r=P_t=P\), Eq.~(\ref{eq:structure_tensor}) reduces to the perfect fluid form used in the standard Tolman Oppenheimer Volkoff construction \cite{Tolman1939,OppenheimerVolkoff1939}.

The dynamics is determined by the Einstein field equations,
\begin{equation}
G^\mu{}_\nu
=
\frac{8\pi G}{c^4}T^\mu{}_\nu.
\label{eq:einstein_equations}
\end{equation}
The temporal component gives
\begin{equation}
\frac{1}{r^2}\frac{d}{dr}
\left[
r\left(1-e^{-2\Lambda}\right)
\right]
=
\frac{8\pi G}{c^4}\epsilon(r).
\label{eq:einstein_temporal_component}
\end{equation}
Using Eq.~(\ref{eq:lambda_mass}), this relation reduces to the mass continuity equation
\begin{equation}
\frac{dm}{dr}
=4\pi r^2\frac{\epsilon}{c^2}.
\label{eq:tov_mass}
\end{equation}
The radial component can be written as
\begin{equation}
-\frac{e^{-2\Lambda}}{r^2}
\left(
2r\frac{d\Phi}{dr}+1
\right)
+\frac{1}{r^2}
=-\frac{8\pi G}{c^4}P_r.
\label{eq:einstein_radial_component}
\end{equation}
Solving Eq.~(\ref{eq:einstein_radial_component}) for the metric potential gives
\begin{equation}
\frac{d\Phi}{dr}
=
\frac{G\left(m+4\pi r^3P_r/c^2\right)}
{c^2r^2\left(1-2Gm/(c^2r)\right)}.
\label{eq:metric_potential_gradient}
\end{equation}
The hydrostatic equation follows from local energy momentum conservation, \(\nabla_\mu T^\mu{}_\nu=0\). For the radial component this condition gives
\begin{equation}
\frac{dP_r}{dr}
=
-\left(\epsilon+P_r\right)\frac{d\Phi}{dr}
+\frac{2\left(P_t-P_r\right)}{r}.
\label{eq:conservation_anisotropic}
\end{equation}
Substitution of Eq.~(\ref{eq:metric_potential_gradient}) into Eq.~(\ref{eq:conservation_anisotropic}) yields the anisotropic generalization of the TOV equation \cite{BowersLiang1974,Herrera1997},
\begin{equation}
\frac{dP_r}{dr}
=
-\frac{G\left(\epsilon+P_r\right)
\left(m+4\pi r^3P_r/c^2\right)}
{c^2r^2\left(1-2Gm/(c^2r)\right)}
+\frac{2\Delta}{r},
\label{eq:anisotropic_tov}
\end{equation}
where
\begin{equation}
\Delta(r)=P_t(r)-P_r(r)
\label{eq:structure_delta}
\end{equation}
is the local anisotropy. The first term in Eq.~(\ref{eq:anisotropic_tov}) is the relativistic gravitational compression written in terms of the radial pressure. The second term is the anisotropic force. A positive \(\Delta\) acts outward, while a negative \(\Delta\) increases the inward effective pressure gradient. When \(\Delta=0\), Eqs.~(\ref{eq:tov_mass}) and (\ref{eq:anisotropic_tov}) reduce to the ordinary isotropic TOV system.

For comparison with the anisotropic solutions, the scalar reference branch is obtained by solving
\begin{equation}
\frac{dP}{dr}
=
-\frac{G\left(\epsilon+P\right)
\left(m+4\pi r^3P/c^2\right)}
{c^2r^2\left(1-2Gm/(c^2r)\right)}.
\label{eq:tov_pressure}
\end{equation}
This reference equation is used only when a single scalar pressure is inserted into the standard structure problem. In the anisotropic calculation the integrated variable is always \(P_r\), while \(P_t\) enters only through \(\Delta\) at the same local energy density.

The magnetic anisotropy is not applied throughout the star. Since the strange core is modeled as isotropic, the anisotropic force is multiplied by a switching function \(S(P_r)\) and the equation actually integrated is
\begin{equation}
\frac{dP_r}{dr}
=
-\frac{G\left(\epsilon+P_r\right)
\left(m+4\pi r^3P_r/c^2\right)}
{c^2r^2\left(1-2Gm/(c^2r)\right)}
+S(P_r)\frac{2\Delta}{r}.
\label{eq:smooth_tov}
\end{equation}
The function \(S(P_r)\) is zero in the strange core and approaches unity inside the crust. The physical distinction used in this work is therefore the binary separation between an isotropic core, \(S=0\), and an anisotropic crust, \(S=1\). A smooth transition was introduced only as a numerical diagnostic to test the efficiency and robustness of the code near the core crust interface.
The localization of the anisotropic force via $S(P_r)$ reflects the physical distinction between the two stellar domains. In the ordinary crust, the pressure is dominated by a degenerate electron gas where the magnetic field directly induces Landau quantization and a measurable magnetization \cite{Strickland_2012}. Conversely, the MIT bag core represents a self-bound quark phase where the internal pressure is dominated by the strong interaction and the bag constant, making it approximately isotropic relative to the crustal magnetic splitting \cite{Glendenning_1995,Menezes_2015}.

For numerical diagnostics, the transition can be written as a smoothstep in logarithmic pressure,
\begin{equation}
S(P_r)=
\begin{cases}
0, & P_r\geq P_{\rm join},\\
3x^2-2x^3, & P_{\rm surf}<P_r<P_{\rm join},\\
1, & P_r\leq P_{\rm surf},
\end{cases}
\label{eq:switch_function}
\end{equation}
with
\begin{equation}
x=
\frac{\ln P_{\rm join}-\ln P_r}
{\ln P_{\rm join}-\ln P_{\rm surf}}.
\label{eq:switch_coordinate}
\end{equation}
Here \(P_{\rm join}\) is the pressure at the core crust interface and \(P_{\rm surf}\) is the lower pressure scale used to complete the transition. This smoothing is not an additional physical layer. It only avoids a sharp numerical change in \(dP_r/dr\) near the interface. The physical model remains the limiting separation \(S=0\) in the strange core and \(S=1\) in the crust. The stellar surface is reached when the radial pressure falls to the prescribed surface value, and the gravitational mass and radius are then \(M=m(R)\) and \(R\), respectively.

\section{Results}
\label{sec:results}

The configurations discussed in this section are built from an isotropic strange matter core and a magnetized ordinary matter crust. The core is described by the MIT bag model used throughout this work and is not assigned a magnetic pressure anisotropy. The magnetic field enters only through the crust equation of state, where the electron gas is quantized and the stress tensor separates the radial and tangential pressure branches. In the stellar structure equations the integrated pressure is \(P_r\), while \(P_t\) contributes through the local anisotropic term \(2(P_t-P_r)/r\). This separation is useful because it keeps the role of the crust explicit. The magnetic field changes the response of the outer layers, whereas the strange matter core fixes the inner boundary condition for the crust and changes the family of equilibrium configurations relative to an ordinary white dwarf.

\begin{figure*}
\includegraphics[width=\textwidth]{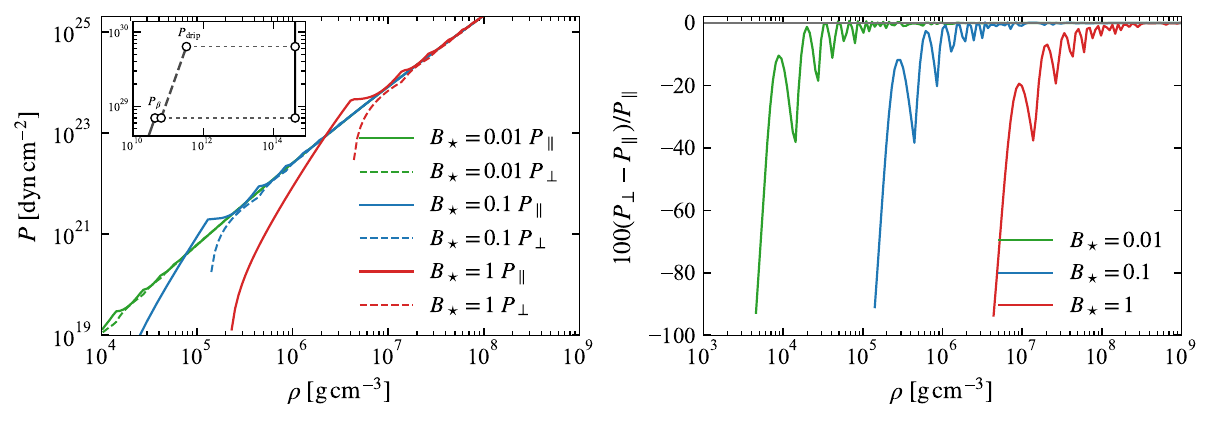}
\caption{\label{fig:eos_core_magnetic_splitting}Core and crust equation of state used in the stellar models. The left panel shows the pressure as a function of mass density in cgs units, including the strange matter core branch and the crust branch up to the neutron drip limit. The vertical markers indicate the beta inverse threshold and the neutron drip threshold. The right panel shows the relative magnetic pressure splitting in the crust, \(100(P_t-P_r)/P_r\), for the tabulated magnetic fields over the interval where both pressure branches are positive.}
\end{figure*}

Figure~\ref{fig:eos_core_magnetic_splitting} establishes the equation of state input used in the following calculations. The carbon crust can be followed as a pure \(^{12}{\rm C}\) sequence only up to the first beta inverse threshold, where electron capture changes the nuclear composition. This type of microscopic stability limit has also been used to constrain massive and magnetized white dwarf sequences with realistic crust ingredients \cite{otoniel_strongly_2019,otoniel_mass_2021,malheiro_relevance_2021}. If the sequence is continued, the crust passes through a more neutron rich composition and reaches the neutron drip density. The neutron drip point marks the limit at which the matter can no longer be described only by nuclei in a degenerate electron background, because free neutrons begin to appear. In the present calculation the crust equation of state is therefore used in two ways. The first construction stops at the beta inverse threshold and represents a pure carbon crust. The second construction follows the tabulated nuclear sequence up to neutron drip and represents the deeper crust allowed before free neutrons appear.

The right panel of Fig.~\ref{fig:eos_core_magnetic_splitting} shows the quantity that enters the anisotropic part of the structure equation. The splitting between \(P_t\) and \(P_r\) is not a smooth rescaling of the pressure. It contains oscillatory features associated with the filling of Landau levels in the magnetized electron gas. Since the tangential branch satisfies \(P_t=P_r-{\cal M}_eB\), the magnetization is transferred directly to \(P_t-P_r\). In the density interval displayed here the splitting is negative for the tabulated crusts, so the term \(2(P_t-P_r)/r\) acts with the sign that steepens the effective radial pressure gradient in the crust. This is a local crust effect in the present model, not a magnetic modification of the strange matter core.
 As illustrated in the right panel of Fig. 1, the pressure splitting is negative ($\Delta = P_t - P_r < 0$) across the crustal density range. In the context of the anisotropic TOV equation (Eq. 49), this negative term increases the inward effective pressure gradient \cite{BowersLiang1974}. Unlike models where positive anisotropy provides additional support against gravity, the magnetic anisotropy here acts as a compressive force that shifts the equilibrium sequences toward significantly smaller radii.

\begin{figure*}
\includegraphics[width=\textwidth]{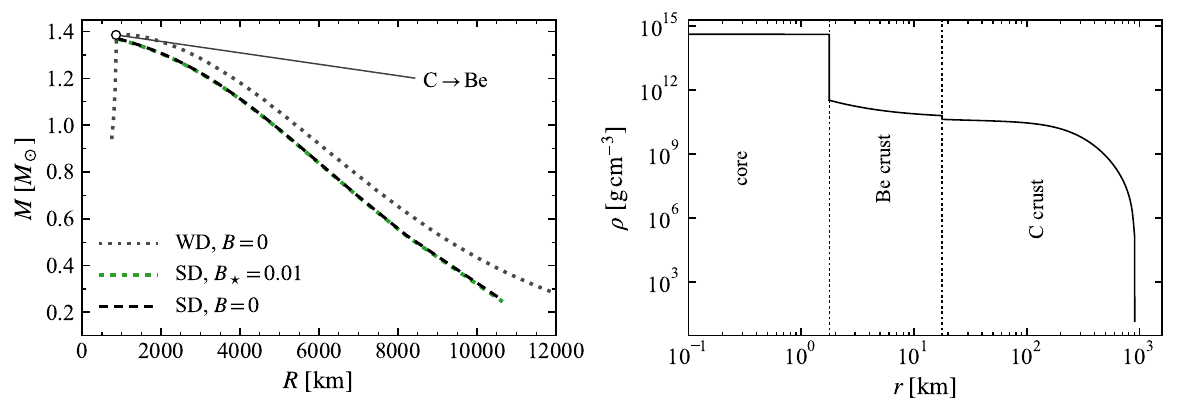}
\caption{\label{fig:wd_sd_branch_profile}Mass radius relation and internal density profile for the carbon sequence extended to neutron drip. The left panel compares an ordinary \(B=0\) white dwarf sequence with selected strange dwarf branches. The right panel shows the density profile of the maximum mass model on the selected \(B=0\) strange dwarf branch, separating the strange matter core, the \(^{12}{\rm Be}\) crust layer, and the outer \(^{12}{\rm C}\) crust.}
\end{figure*}

The effect of the nuclear sequence on the stellar family is shown in Fig.~\ref{fig:wd_sd_branch_profile}. The ordinary \(B=0\) white dwarf curve is sensitive to the change in composition that occurs as the sequence approaches neutron drip. The reduction of \(Z/A\) lowers the electron pressure support at a given mass density and softens the crust equation of state. In the mass radius plane this appears as a departure from the usual white dwarf branch after the composition has changed. The strange dwarf sequence behaves differently because the central region is a self bound strange matter core rather than ordinary ion and electron matter. The core supplies the inner boundary for the crust and allows equilibrium configurations to be followed through a density interval in which the ordinary white dwarf sequence does not provide the same continuous branch.

For the strange dwarf curves in Fig.~\ref{fig:wd_sd_branch_profile}, the plotted segment is the branch selected from the maximum radius point to the maximum mass point of the fixed \(P_{\rm drip}\) sequence. This selection gives a working interval with positive \(dM/d\rho_c\) in the sampled family, but it should not be read as a full radial stability proof.
For \(B=0\), the selected branch spans \(4.1274\times10^{14}\leq\rho_c\leq4.1705\times10^{14}\,{\rm g\,cm^{-3}}\), or equivalently \(231.53\leq\epsilon_c\leq233.95\,{\rm MeV\,fm^{-3}}\). Along this branch the mass increases from \(0.262\,M_\odot\) at \(R=1.0555\times10^4\,{\rm km}\) to \(1.369\,M_\odot\) at \(R=903\,{\rm km}\). For \(B_\star=0.01\), the same central density interval gives a branch from \(0.245\,M_\odot\) at \(R=1.0652\times10^4\,{\rm km}\) to \(1.367\,M_\odot\) at \(R=902\,{\rm km}\). The right panel gives the density profile of the maximum mass \(B=0\) model on this branch. The model has a compact strange matter core of radius \(1.77\,{\rm km}\), followed by a \(^{12}{\rm Be}\) layer extending to about \(17.7\,{\rm km}\), and an outer \(^{12}{\rm C}\) crust that carries most of the stellar radius.

\begin{figure}
\includegraphics[width=\columnwidth]{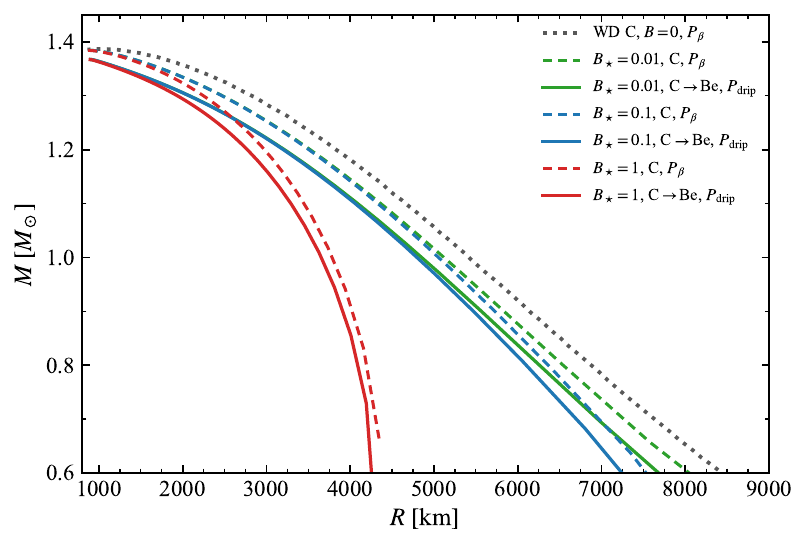}
\caption{\label{fig:c12_fixed_pcc_mr}Mass radius relation for carbon strange dwarfs with physically fixed core crust transition pressures. Dashed curves use a pure \(^{12}{\rm C}\) crust stopped at the beta inverse threshold, \(P_{\rm cc}=P_\beta\). Solid curves use the \(^{12}{\rm C}\rightarrow^{12}{\rm Be}\) sequence extended to neutron drip, \(P_{\rm cc}=P_{\rm drip}\). The dotted curve is the ordinary \(B=0\) white dwarf comparison.}
\end{figure}

Figure~\ref{fig:c12_fixed_pcc_mr} is the central comparison for the carbon crust. The dashed curves use the pure carbon crust only up to the beta inverse threshold. The solid curves instead use the full \(^{12}{\rm C}\rightarrow^{12}{\rm Be}\) sequence up to neutron drip. These are different physical choices for the core crust transition pressure, not only different plotting styles. The pure carbon curves use \(P_{\rm cc}=P_\beta\), while the full sequence uses \(P_{\rm cc}=P_{\rm drip}\). At fixed magnetic field, extending the crust to neutron drip shifts the selected strange dwarf branch toward smaller radii at comparable masses. This behavior follows from the softer compositionally evolved crust attached to the strange matter core.

The same figure also shows the effect of the magnetic anisotropy. For both transition choices, increasing \(B_\star\) moves the selected branch to smaller radii over the plotted interval. The displacement is more visible at lower masses, where the crust occupies a larger fraction of the stellar radius. Near the maximum mass the curves remain close because the compact strange matter core controls more of the structure. In the carbon sequence extended to neutron drip, the maximum mass changes only moderately across the fields shown: \(1.3675\,M_\odot\) for \(B_\star=0.01\), \(1.3675\,M_\odot\) for \(B_\star=0.1\), and \(1.3672\,M_\odot\) for \(B_\star=1\). The corresponding radii at the maximum remain near \(900\,{\rm km}\). The main magnetic effect in this mass window is therefore a reduction of the radius along the branch, accompanied by a small decrease of the maximum mass for the largest field considered here.

\begin{figure*}
\includegraphics[width=\textwidth]{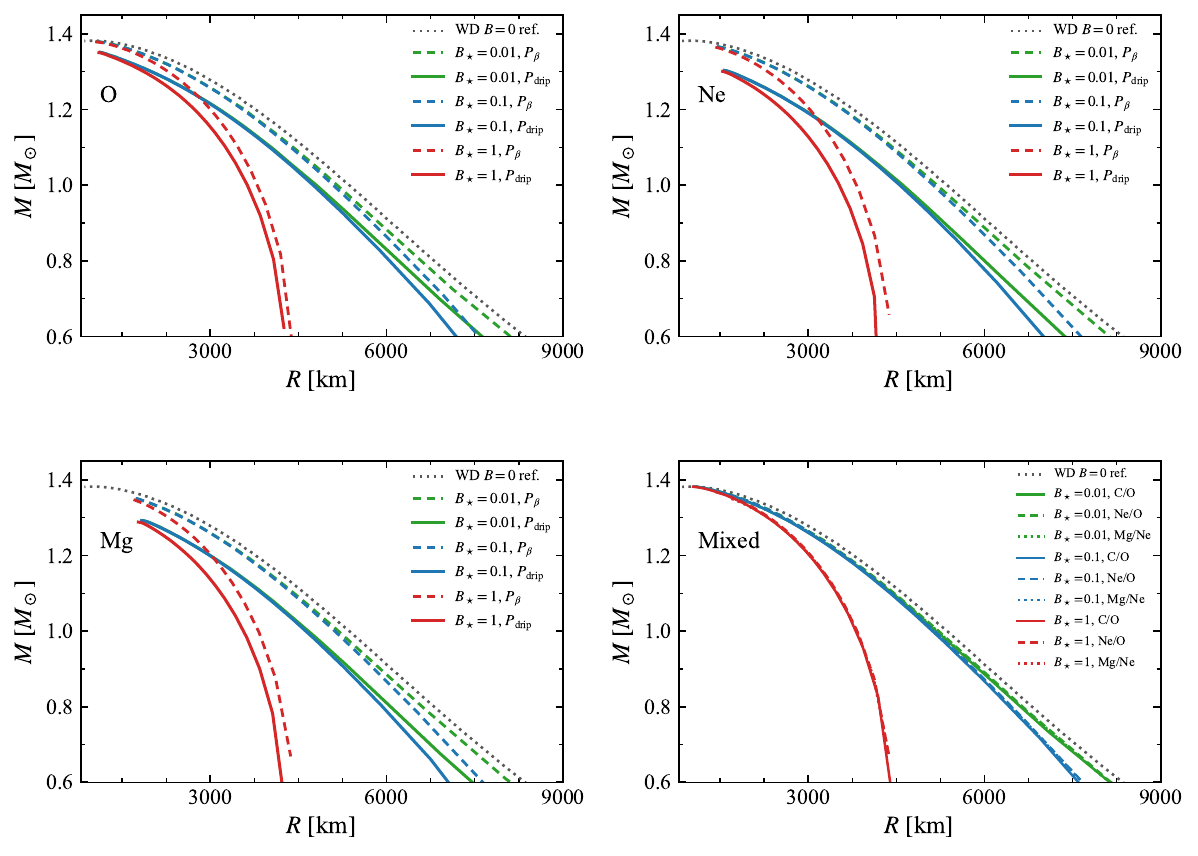}
\caption{\label{fig:elements_mixed_fixed_pcc_mr}Mass radius relation for O, Ne, Mg, and mixed crusts. In the O, Ne, and Mg panels, dashed curves stop at the first beta inverse threshold and solid curves follow the sequence to neutron drip. The mixed panel shows 50/50 C/O, Ne/O, and Mg/Ne crusts up to the first beta inverse threshold. The dotted curve is the available ordinary \(B=0\) white dwarf reference used for comparison.}
\end{figure*}

Figure~\ref{fig:elements_mixed_fixed_pcc_mr} extends the same construction to heavier nuclei and to binary mixtures. The O, Ne, and Mg panels repeat the comparison between a crust stopped at the first beta inverse threshold and a sequence continued to neutron drip. The detailed transitions differ from the carbon case because each initial nucleus follows its own capture sequence, but the effect on the selected strange dwarf branch is similar. The curves that reach neutron drip are generally more compact than the corresponding beta limited curves, and the magnetic field again shifts the branches toward smaller radii. The reduction in maximum mass with increasing \(B_\star\) remains modest in the plotted range, while the change in radius is clearer along the lower mass part of each branch.

The mixed panel in Fig.~\ref{fig:elements_mixed_fixed_pcc_mr} should be read separately from the pure O, Ne, and Mg panels. The 50/50 C/O, Ne/O, and Mg/Ne tables are stopped at the first beta inverse instability, so they test the sensitivity to mixed composition before a complete neutron drip sequence is included. They nevertheless provide a useful comparison because the ordering with \(B_\star\) remains consistent with the pure composition panels. The ordinary \(B=0\) white dwarf curve in this figure is the available local reference curve and is not an element specific \(B=0\) sequence for each panel. With this limitation, the set of panels indicates that the compactification found for carbon is also present for heavier crust compositions in the same modeling framework.

The compactification at fixed stellar mass is summarized in Table~\ref{tab:sd_one_solar_mass_core}. All entries are evaluated at \(M=1\,M_\odot\), obtained by interpolation along the selected strange dwarf branches, and the compositions are ordered from the lightest parent nucleus to the heaviest one. This choice removes the small differences between the maximum mass points and makes the radial effect easier to compare across magnetic fields and compositions. For the carbon sequence, increasing the field from \(B_\star=0.01\) to \(B_\star=1\) reduces the radius from \(5118\,{\rm km}\) to \(3783\,{\rm km}\) for the beta limited branch, and from \(4851\,{\rm km}\) to \(3653\,{\rm km}\) for the neutron drip branch. At the same mass, the neutron drip branches also have larger strange cores than the beta limited branches, because the transition pressure is deeper in the crust sequence.

\begin{table*}
\caption{\label{tab:sd_one_solar_mass_core}Interpolated properties of selected strange dwarf branches at fixed stellar mass \(M=1\,M_\odot\). The pure entries are ordered from carbon to magnesium. The columns \(M_{\rm core}\), \(R_{\rm core}\), and \(\rho_c\) refer to the same interpolated model.}
\begin{ruledtabular}
\begin{tabular}{lcccccc}
Crust model & Limit & \(B_\star\) & \(R\,[{\rm km}]\) & \(M_{\rm core}/M_\odot\) & \(R_{\rm core}\,[{\rm km}]\) & \(\rho_c\,[{\rm g\,cm^{-3}}]\)\\
\colrule
\multicolumn{7}{c}{Pure branches}\\
\(^{12}{\rm C}\) & \(P_\beta\) & 0.01 & 5118 & \(1.02\times10^{-2}\) & 2.273 & \(4.144\times10^{14}\)\\
\(^{12}{\rm C}\) & \(P_\beta\) & 0.1 & 5053 & \(1.02\times10^{-2}\) & 2.273 & \(4.144\times10^{14}\)\\
\(^{12}{\rm C}\) & \(P_\beta\) & 1 & 3783 & \(9.66\times10^{-3}\) & 2.233 & \(4.143\times10^{14}\)\\
\(^{12}{\rm C}\rightarrow{}^{12}{\rm Be}\) & \(P_{\rm drip}\) & 0.01 & 4851 & \(1.94\times10^{-2}\) & 2.812 & \(4.167\times10^{14}\)\\
\(^{12}{\rm C}\rightarrow{}^{12}{\rm Be}\) & \(P_{\rm drip}\) & 0.1 & 4796 & \(1.93\times10^{-2}\) & 2.812 & \(4.167\times10^{14}\)\\
\(^{12}{\rm C}\rightarrow{}^{12}{\rm Be}\) & \(P_{\rm drip}\) & 1 & 3653 & \(1.87\times10^{-2}\) & 2.782 & \(4.166\times10^{14}\)\\
\(^{16}{\rm O}\) & \(P_\beta\) & 0.01 & 5158 & \(6.86\times10^{-3}\) & 1.992 & \(4.134\times10^{14}\)\\
\(^{16}{\rm O}\) & \(P_\beta\) & 0.1 & 5090 & \(6.85\times10^{-3}\) & 1.991 & \(4.134\times10^{14}\)\\
\(^{16}{\rm O}\) & \(P_\beta\) & 1 & 3798 & \(6.37\times10^{-3}\) & 1.944 & \(4.133\times10^{14}\)\\
\(^{16}{\rm O}\rightarrow{}^{16}{\rm C}\) & \(P_{\rm drip}\) & 0.01 & 4805 & \(1.92\times10^{-2}\) & 2.803 & \(4.167\times10^{14}\)\\
\(^{16}{\rm O}\rightarrow{}^{16}{\rm C}\) & \(P_{\rm drip}\) & 0.1 & 4750 & \(1.92\times10^{-2}\) & 2.803 & \(4.167\times10^{14}\)\\
\(^{16}{\rm O}\rightarrow{}^{16}{\rm C}\) & \(P_{\rm drip}\) & 1 & 3631 & \(1.85\times10^{-2}\) & 2.773 & \(4.165\times10^{14}\)\\
\(^{20}{\rm Ne}\) & \(P_\beta\) & 0.01 & 5193 & \(3.57\times10^{-3}\) & 1.604 & \(4.123\times10^{14}\)\\
\(^{20}{\rm Ne}\) & \(P_\beta\) & 0.1 & 5128 & \(3.56\times10^{-3}\) & 1.603 & \(4.123\times10^{14}\)\\
\(^{20}{\rm Ne}\) & \(P_\beta\) & 1 & 3823 & \(3.17\times10^{-3}\) & 1.541 & \(4.121\times10^{14}\)\\
\(^{20}{\rm Ne}\rightarrow{}^{20}{\rm O}\rightarrow{}^{20}{\rm C}\) & \(P_{\rm drip}\) & 0.01 & 4612 & \(2.37\times10^{-2}\) & 3.009 & \(4.177\times10^{14}\)\\
\(^{20}{\rm Ne}\rightarrow{}^{20}{\rm O}\rightarrow{}^{20}{\rm C}\) & \(P_{\rm drip}\) & 0.1 & 4561 & \(2.37\times10^{-2}\) & 3.009 & \(4.177\times10^{14}\)\\
\(^{20}{\rm Ne}\rightarrow{}^{20}{\rm O}\rightarrow{}^{20}{\rm C}\) & \(P_{\rm drip}\) & 1 & 3531 & \(2.31\times10^{-2}\) & 2.982 & \(4.175\times10^{14}\)\\
\(^{24}{\rm Mg}\) & \(P_\beta\) & 0.01 & 5178 & \(2.34\times10^{-3}\) & 1.394 & \(4.118\times10^{14}\)\\
\(^{24}{\rm Mg}\) & \(P_\beta\) & 0.1 & 5116 & \(2.34\times10^{-3}\) & 1.393 & \(4.118\times10^{14}\)\\
\(^{24}{\rm Mg}\) & \(P_\beta\) & 1 & 3819 & \(2.00\times10^{-3}\) & 1.321 & \(4.116\times10^{14}\)\\
\(^{24}{\rm Mg}\rightarrow{}^{24}{\rm Ne}\rightarrow{}^{24}{\rm O}\) & \(P_{\rm drip}\) & 0.01 & 4667 & \(2.00\times10^{-2}\) & 2.844 & \(4.169\times10^{14}\)\\
\(^{24}{\rm Mg}\rightarrow{}^{24}{\rm Ne}\rightarrow{}^{24}{\rm O}\) & \(P_{\rm drip}\) & 0.1 & 4624 & \(2.00\times10^{-2}\) & 2.843 & \(4.169\times10^{14}\)\\
\(^{24}{\rm Mg}\rightarrow{}^{24}{\rm Ne}\rightarrow{}^{24}{\rm O}\) & \(P_{\rm drip}\) & 1 & 3564 & \(1.94\times10^{-2}\) & 2.814 & \(4.167\times10^{14}\)\\
\colrule
\multicolumn{7}{c}{Binary \(50/50\) mixed branches}\\
\(^{12}{\rm C}/^{16}{\rm O}\) & \(P_\beta\) & 0.01 & 5178 & \(6.87\times10^{-3}\) & 1.994 & \(4.134\times10^{14}\)\\
\(^{12}{\rm C}/^{16}{\rm O}\) & \(P_\beta\) & 0.1 & 5112 & \(6.86\times10^{-3}\) & 1.993 & \(4.134\times10^{14}\)\\
\(^{12}{\rm C}/^{16}{\rm O}\) & \(P_\beta\) & 1 & 3810 & \(6.39\times10^{-3}\) & 1.946 & \(4.133\times10^{14}\)\\
\(^{20}{\rm Ne}/^{16}{\rm O}\) & \(P_\beta\) & 0.01 & 5220 & \(3.58\times10^{-3}\) & 1.605 & \(4.123\times10^{14}\)\\
\(^{20}{\rm Ne}/^{16}{\rm O}\) & \(P_\beta\) & 0.1 & 5152 & \(3.57\times10^{-3}\) & 1.604 & \(4.123\times10^{14}\)\\
\(^{20}{\rm Ne}/^{16}{\rm O}\) & \(P_\beta\) & 1 & 3834 & \(3.18\times10^{-3}\) & 1.543 & \(4.121\times10^{14}\)\\
\(^{24}{\rm Mg}/^{20}{\rm Ne}\) & \(P_\beta\) & 0.01 & 5199 & \(2.35\times10^{-3}\) & 1.395 & \(4.118\times10^{14}\)\\
\(^{24}{\rm Mg}/^{20}{\rm Ne}\) & \(P_\beta\) & 0.1 & 5137 & \(2.34\times10^{-3}\) & 1.394 & \(4.118\times10^{14}\)\\
\(^{24}{\rm Mg}/^{20}{\rm Ne}\) & \(P_\beta\) & 1 & 3829 & \(2.00\times10^{-3}\) & 1.323 & \(4.116\times10^{14}\)\\
\end{tabular}
\end{ruledtabular}
\end{table*}

\begin{figure}
\includegraphics[width=\columnwidth]{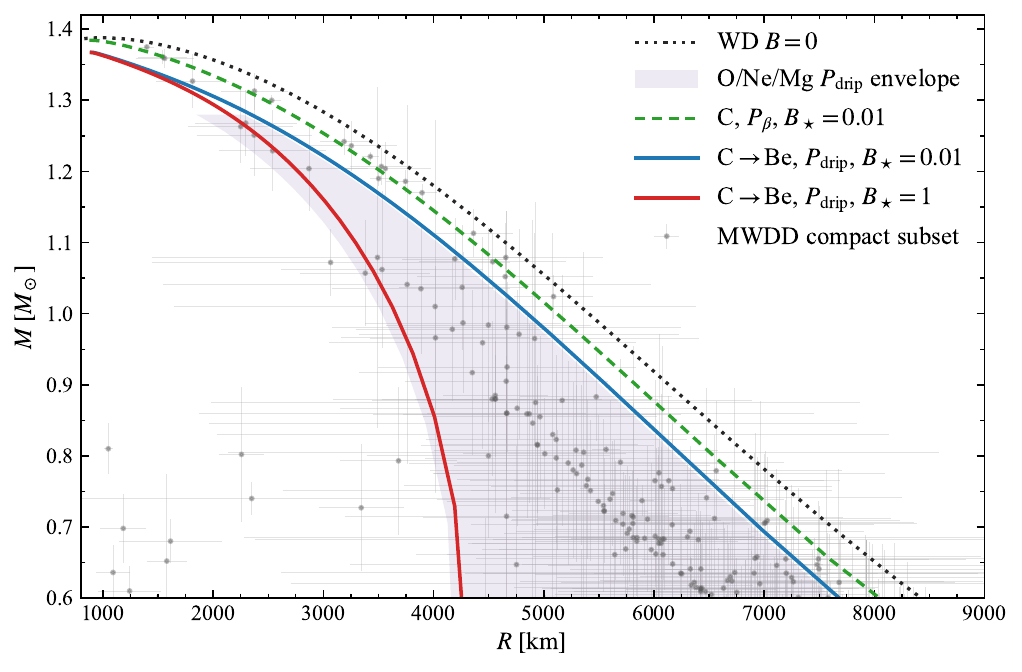}
\caption{\label{fig:observational_context_mr}Mass radius context for selected strange dwarf branches and a compact subset of objects from the Montreal White Dwarf Database. The catalog radius was inferred from the tabulated mass and surface gravity through \(R=(GM/g)^{1/2}\), with \(g=10^{\log g}\). The comparison is intended only as an observational reference plane for the calculated sequences.}
\end{figure}

Taken together, the five figures separate the ingredients that control the result. The equation of state figure identifies the beta inverse and neutron drip thresholds and shows the pressure splitting that produces the crust anisotropy. The white dwarf comparison shows why the composition sequence changes the ordinary mass radius curve and why a strange matter core permits a selected branch to be followed through the same density interval. The carbon figure shows the combined effect of fixing \(P_{\rm cc}\) at the physical transition pressure and increasing \(B_\star\). The O, Ne, Mg, and mixed composition figure shows that the same trend persists when the crust composition is changed. 

Finally, Fig.~\ref{fig:observational_context_mr} places representative branches in the same mass radius plane as a compact subset of objects exported from the Montreal White Dwarf Database \cite{Dufour2017MWDD,MWDDOnline}. Although these objects are catalogued as white dwarfs, the small strange matter core and the ordinary outer crust of a strange dwarf can make the two classes difficult to separate from global mass and radius information alone. The plotted radii are inferred from the catalog mass and \(\log g\) through the Newtonian surface gravity relation \(g=GM/R^2\). In this sense, the comparison suggests that the most compact white dwarf candidates are a useful observational region to examine with more detailed modeling, especially when magnetic anisotropy is included because it shifts the calculated branches toward smaller radii. This is not a source by source identification, but a guide for where possible strange dwarf candidates could be searched for with additional spectroscopic, magnetic, and evolutionary constraints. A radial mode calculation with the appropriate core crust boundary conditions is still required before the selected branches can be called dynamically stable in the strict sense.
The comparison with the Montreal White Dwarf Database highlights that the localized magnetic anisotropy moves the strange dwarf branches into the high-gravity region ($\log g$) of the mass-radius plane. This suggests that the most compact white dwarf candidates are the ideal observational targets for testing strange dwarf scenarios, as their radii are often difficult to reconcile with standard isotropic electron-degenerate models \cite{Mathews_2006, otoniel_strongly_2019}.

\section{Discussion and conclusions}
\label{sec:conclusions}

We have presented equilibrium sequences for strange dwarfs composed of a self bound strange matter core described by the MIT bag model with \(B_{\rm bag}^{1/4}=145\,{\rm MeV}\), surrounded by an ordinary matter crust. Within this setup, the magnetic field is introduced through the crust EOS and produces different radial and tangential pressure components. The main point of the calculation is that this pressure anisotropy should enter the stellar structure as a stress contribution in the crust, rather than as a replacement of the pressure by one scalar branch of the EOS. Scalar integrations using either \(P_r\) or \(P_t\) are useful controls because they show the direct effect of choosing one pressure component in an otherwise ordinary TOV problem. They do not, however, include the force term associated with the anisotropic stress tensor. Once the hydrostatic equation is solved with \(P_r\) as the integrated pressure and \(2(P_t-P_r)/r\) as a localized crust contribution, the mass radius relation changes through the radial force balance itself.

The composition of the crust also plays a direct role. A pure crust stopped at the first beta inverse threshold and a sequence continued to neutron drip correspond to different physical choices for the core crust transition pressure. The first case describes a crust limited by the onset of beta instability in the parent nucleus, while the second follows the capture sequence into a stratified crust until the neutron drip point is reached. In the mass radius sequences, the neutron drip branches are more compact than the corresponding beta limited branches. This behavior follows from the softer compositionally evolved crust, and it is seen not only for carbon but also in the O, Ne, and Mg sequences considered here.

The difference between an ordinary white dwarf sequence and a strange dwarf sequence is important for interpreting this result. In an ordinary white dwarf, the change in composition near the high density end of the crust modifies the electron pressure support and can drive the mass radius curve away from the usual stable branch. In a strange dwarf, the ordinary crust is attached to a self bound strange matter core. This core provides an inner boundary that permits equilibrium configurations to be followed through a selected density interval in which the ordinary white dwarf sequence does not provide the same continuous branch. This statement concerns the equilibrium families obtained in the present TOV treatment and should not be read as a proof of dynamical stability.

For the magnetized EOS tables used here, the relevant crust intervals have \(P_t-P_r<0\). The anisotropic term therefore steepens the effective pressure gradient in the crust and shifts the selected strange dwarf branches toward smaller radii as \(B_\star\) increases. This compactification appears in the carbon sequence and remains visible in the heavier crust compositions and in the mixed beta limited branches. The effect is a crustal anisotropy effect in the present model. The strange matter core is kept isotropic, and the calculation does not make a general statement about stars in which the magnetic field and its stresses are distributed throughout the dense interior.

The fixed mass comparison in Table~\ref{tab:sd_one_solar_mass_core} gives a compact summary of the same trend. Evaluating the models at \(M=1\,M_\odot\) separates the radial response from small differences between maximum mass points. At this mass, increasing \(B_\star\) reduces the stellar radius across the compositions shown, while the neutron drip branches generally have larger strange cores than the beta limited branches because the core crust transition occurs at a higher pressure. The table therefore connects the global compactification to the internal partition between the strange core and the ordinary crust.

The observational comparison in Fig.~\ref{fig:observational_context_mr} should be interpreted in the same cautious way. The objects plotted from the Montreal White Dwarf Database are catalogued as white dwarfs, but a strange dwarf with a small strange matter core and an ordinary outer crust can be difficult to distinguish from a white dwarf using only global mass and radius information. The figure therefore does not identify any source as a strange dwarf. It indicates that the compact part of the observed white dwarf mass radius plane is a relevant region for future tests of this scenario, especially because magnetic anisotropy moves the calculated branches toward smaller radii. Any candidate level interpretation would require additional information, including atmospheric composition, magnetic field estimates, cooling history, and a consistent evolutionary picture.

Several limitations remain. The stellar structure is described by an effective spherical anisotropic equation, not by a self consistent axisymmetric magnetic equilibrium. The magnetic anisotropy is applied only in the ordinary matter crust, and the strange quark matter core is kept isotropic. The Maxwell stress is not included as an independent global source in the stellar equations, consistent with the EOS implementation adopted here. The mixed composition models are currently stopped at the first beta inverse threshold, while complete mixed sequences to neutron drip still need to be constructed. The smooth core crust transition used in the numerical code should also be regarded as an interface diagnostic rather than a separate physical layer.

The next steps are therefore clear. A radial stability calculation with boundary conditions appropriate to the strange matter core and ordinary matter crust interface is required before the selected branches can be called dynamically stable. The crust tables should be extended to more complete mixed compositions and to a broader survey of magnetic field strengths. The effective spherical treatment should be compared with axisymmetric magnetic models in which the field geometry and deformation are solved explicitly. On the observational side, the compact white dwarf region highlighted by the MWDD comparison provides a useful target for more detailed searches, but only when mass, radius, magnetic information, spectroscopy, and evolutionary constraints are analyzed together.

In summary, the present results show how crust composition, core crust matching, and magnetic pressure anisotropy act together in strange dwarf models. Extending the crust to neutron drip makes the selected branches more compact than the beta limited case, while the anisotropic crust force produces an additional shift toward smaller radii as the magnetic field increases. Within the assumptions of this work, magnetic pressure anisotropy is therefore best viewed as a localized stress of the ordinary matter crust, coupled to an isotropic strange matter core, rather than as a scalar modification of the pressure throughout the star.

\begin{acknowledgments}
This swork was financed in part by the Coordenação de Aperfeiçoamento de Pessoal de Nível Superior - Brasil (CAPES) - Finance Code 001, and Fundação de Amparo à Pesquisa e Inovação do Espírito Santo (FAPES). J.G.C. is
grateful for the support of FAPES (1020/2022, 1081/2022, 976/2022,
332/2023, 1514/2025), CNPq (311758/2021-5, 306018/2025-0), and FAPESP (grant
No. 2021/01089-1).
E.O. acknowledges support from the Funda\c{c}\~ao Cearense de Apoio ao Desenvolvimento Cient\'ifico e Tecnol\'ogico (FUNCAP) through grant BP6-0241-00335.01.00/25.
\end{acknowledgments}

\bibliographystyle{apsrev4-2}
\bibliography{wdmdr}

\end{document}